\documentclass[12pt,amssymb]{iopart}
\usepackage{graphicx,epsfig}
\usepackage{graphicx}
\usepackage{dcolumn}
\usepackage{bm}
\usepackage{epsfig}
\usepackage{graphics}
\usepackage{amsfonts}

\begin{document}
\title[Resonant states  in an attractive one dimensional
cusp ...] {Resonant states  in an attractive one dimensional cusp
potential}
\author{
V\'{\i}ctor M Villalba\footnote[1]{E-mail: villalba@ivic.ve}
\footnote[2]{Alexander von Humboldt Fellow}, Luis A.
Gonz\'alez-D\'{\i}az}
\address{Centro de F\'{\i}sica IVIC Apdo 21827, Caracas 1020A,
Venezuela}

\begin{abstract}
We solve the two-component Dirac equation in the presence of a
spatially one dimensional symmetric attractive cusp potential. The
components of the spinor solution are expressed in terms of
Whittaker functions. We compute the bound states solutions and show
that, as the potential amplitude increases, the lowest energy state
sinks into the Dirac sea becoming a resonance. We characterize and
compute the lifetime of the resonant state with the help of the
phase shift and the Breit-Wigner relation. We discuss the limit when
the cusp potential reduces to a delta point interaction.
\end{abstract}


\maketitle

\section{Introduction}

Spontaneous particle production in the presence of strong electric
fields is perhaps one of the most interesting phenomena associated
with the charged quantum vacuum \cite{Greiner,Rafelski}. The study
of supercritical effects and resonant particle production by strong
Coulomb-like potentials  dates back to the pioneering works of
Pieper and Greiner \cite{Pieper} and Gershtein and Zeldovich
\cite{Gershtein} where it was shown that spontaneous positron
production was possible when two heavy bare nuclei with total charge
larger than some critical value $Z_{c}$ collided with each other.
The critical $Z_{c}$ is the value for which the $1S$ state of the
hydrogenlike atom with total charge Z has energy $E=-m$.  The idea
behind supercriticality lies in spontaneous positron emission
induced by the presence of very strong attractive vector potentials.
The presence of a strong electric field induces the energy level of
an unoccupied bound state to sink into the negative energy
continuum, i.e, an electron of the Dirac sea is trapped by the
potential, leaving a positron that escapes to infinity.  The
electric field responsible for supercritical effects should be
stronger than $2m_{e}c^{2}$, which is the value of the gap between
the negative and positive energy continua.  Such strong electric
fields could be produced in heavy-ion collisions
\cite{Greiner,Greiner2,Reinhardt}

In the last years the one-dimensional Dirac equation in an external potential $V$
has been studied in connection to the  Levinson theorem in relativistic
quantum mechanics \cite{Lin,Calogeracos,Ma}. The interest in this problem
can be found in their applications in semiconductor physics, field theory and
one-dimensional QED among others. One-dimensional Dirac electrons can be
considered as three-dimensional problems when $V$ depends only
on one space variable \cite{Lin}.   The relation between supercritical states and
the Levinson theorem in the one-dimensional Dirac has been established by
Ma \textit{et al} \cite{Ma}, and Calogeracos and Dombey \cite{Calogeracos}.

Recently the relation between transmission resonances and
supercriticality effects have been studied in a one-dimensional
barrier \cite{Dombey} and in a Woods-Saxon potential
\cite{Kennedy,Dosch},  which is a smoothed form of a square well.
Here the Dirac equation presents half-bound states with the same
asymptotic behavior of those obtained with the potential barrier.
Supercritical states with $E=-m$  have been also studied in a
one-dimensional cusp potential \cite{villalba,Jiang} and in a class
of short-range potentials \cite{KD:04}.

In the present article we discuss the problem of computing the
phenomenon of resonant particle creation in a one-dimensional
symmetric cusp potential of the form

\begin{equation}
eU(x)=-V\exp (-\left| x\right| /a).  \label{A1}
\end{equation}
where $V>0$,  $a>0$ and $e$ is the charge of the particle.

The potential (\ref{A1}) is an asymptotically vanishing potential
for large values of the space variable $x$. The parameter $V$ shows
the depth of the well. The constant $a$ determines the shape of the
potential. The expression (\ref{A1}) can be regarded as a screened
one dimensional Coulomb potential \cite{Dominguezadame}.

It is well known that a vectorial delta potential is strong enough
to pull the bound state into the negative energy continuum $E=-m$
\cite{FD:89,Nogami}, nevertheless this supercritical state does not
evolve to a real resonant state. It is of interest to investigate
the resonant behavior of the lowest energy states in a potential
exhibiting a delta point interaction as asymptotic limit. The cusp
potential (\ref{A1}) reduces to an attractive vectorial delta
interaction of strength $g_v$ in the limit $2Va\rightarrow g_v$, as
$a\rightarrow 0$. The attractive potential (\ref{A1})  does not
possess compact support and, in contrast to the delta potential
case, it can exhibit more than one bound state.

It is the purpose of the present article to show that the cusp
potential (\ref{A1}) is strong enough to sink the lowest electron
bound state into the negative continuum to create a resonant state.
In order to calculate the process of embedding  of the bound state
into the negative energy continuum we construct the Jost solutions
and their corresponding Jost functions \cite{Galindo,Thaller}. We
estimate, with the help of the phase shift and the Wigner time-delay
\cite{Newton,Goldberger}, the mean life of the resonance and how it
depends on the shape and strength of the potential.

The article is structured as follows. In section 2 we solve the 1+1
Dirac equation in the potential (\ref{A1}) and derive the equation
governing bound states, energy resonances and supercritical states.
In section 3 we compute, with the help of the Jost functions, the
transmission and reflection amplitudes and the phase shifts. We
derive the condition for energy resonances and estimate the mean
life of the antiparticle state.  Finally, in section 4 we summarize
our results.

\section{Solution of the Dirac equation and energy resonances}

In two dimensions, the Dirac equation  in the presence of the
spatially dependent electric field (\ref{A1}) can be written as
\cite{villalba}

\begin{equation}
\left\{ \gamma ^{0}(\partial _{t}-iV\exp (-\left| x\right|
/a))+\gamma ^{1}\partial _{x}+m\right\} \Psi =0  \label{Dirac}
\end{equation}
where the $\gamma ^{\mu }$ matrices satisfy the commutation relation $%
\left\{ \gamma ^{\mu },\gamma ^{\nu }\right\} =2\eta ^{\mu \nu }$
and the metric $\eta ^{\mu \nu }$ has the signature $(-,+).$ Here
and along the paper, we adopt the natural units $\hbar =c=1.$

Since we are working in 1+1 dimensions, it is possible to choose the
following representation of the Dirac matrices
\begin{equation}
\gamma ^{0}=i\sigma ^{2}=\left(
\begin{array}{cc}
0 & 1 \\
-1 & 0
\end{array}
\right) ,\ \gamma ^{1}=\sigma ^{1}=\left(
\begin{array}{cc}
0 & 1 \\
1 & 0
\end{array}
\right)  \label{rep}
\end{equation}
Taking into account that the potential (\ref{A1}) does not depend on
time, we can write the spinor $\Psi $ as follows
\begin{equation}
\Psi =\left(
\begin{array}{c}
\Psi _{1}(x) \\
\Psi _{2}(x)
\end{array}
\right) \exp (-iEt)  \label{Psi}
\end{equation}
Substituting the gamma  $\gamma^{\mu}$ matrices (\ref{rep}) into the
Dirac equation (\ref {Dirac}), we obtain the following system of
coupled differential  equations
\begin{equation}
\left( \frac{d}{dx}+i(V\exp (-\left| x\right| /a)+E)\right) \Psi
_{1}+m\Psi _{2}=0  \label{a1}
\end{equation}
\begin{equation}
\left( \frac{d}{dx}-i(V\exp (-\left| x\right| /a)+E)\right) \Psi
_{2}+m\Psi _{1}=0  \label{a2}
\end{equation}
The spinor solution of the system (\ref{a1})-(\ref{a2}), exhibiting
a regular behavior as $x\rightarrow +\infty$, can be expressed in
terms of the Whittaker function $M_{k,\mu }(y)$ \cite {Abramowitz}
as follows

\begin{equation}
\Psi _{right}(y)=\left(\begin{array}{c}\Psi _{1} \\ \Psi
_{2}\end{array}\right)=b_{1} \left(\begin{array}{c}
\frac{1/2+\mu +k}{ma}y^{-1/2}M_{k+1,\mu }(y) \\
y^{-1/2}M_{k,\mu }(y)
\end{array}
\right)   \label{inco}
\end{equation}
where $b_{1}$ is a constant, $y=-2iaV\exp (-x/a)$ and
\begin{equation}
k=iEa-1/2,\ \mu =ia\sqrt{E^{2}-m^{2}}  \label{ka}
\end{equation}
For negative values of $x$, we have that the solution of the system
(\ref{a1})-(\ref{a2}) showing a regular behavior as $x\rightarrow
-\infty$ can be expressed in terms of the Whittaker functions
$M_{k,\mu}(y)$ as

\begin{equation}
\Psi _{left}(\bar{y})=\left(\begin{array}{c}\Psi _{1} \\ \Psi
_{2}\end{array}\right)=c_{1}\left(
\begin{array}{c}
\bar{y}^{-1/2}M_{k,\mu }(\bar{y}) \\
-\frac{\frac{1}{2}+\mu +k}{ma}\bar{y}^{-1/2}M_{k+1,\mu }(\bar{y})
\end{array}
\right),   \label{inco2}
\end{equation}
where $c_{1}$ is a constant and $\bar{y}=-2aiV\exp(x/a)$

Imposing the continuity of the spinor solution $\Psi$ at $x=0$, and
demanding  the existence of non-trivial values for $c_{1}$ and
$b_{1}$, we obtain that resonances and  bound levels satisfy the
energy condition:

\begin{equation}
\left( \frac{1}{2}+\mu +k\right) ^{2}M_{k+1,\mu
}^{2}(-2ieaV)+m^{2}a^{2}M_{k,\mu }^{2}(-2ieaV)=0 \label{energy}
\end{equation}
Given the values of the strength $V$ and the shape parameter $a$,
the expression (\ref{energy}) exhibits solutions with $E$ real,
showing that the cusp potential (\ref{A1}) is able to bind
particles, a property that, for instance, is absent in the pure
one-dimensional Coulomb potential.  The number of energy levels in
the cusp potential decreases as the shape parameter decreases.

As the potential depth increases for large values of $V$,  the
energy eigenvalues of the bound states decrease. When
the lowest energy bound state reaches the value $E=-m$, this
level merges with the negative energy continuum and the potential
becomes
supercritical. For $E=-m,$ we have that $\mu =0$ and, using Eq.
(\ref{energy}), we obtain the condition for supercriticality

\begin{equation}
M_{k,0}(-2ieaV_c)M_{k,0}(-2ieaV_c)-M_{k+1,0}(-2ieaV_c)M_{k+1,0}(-2ieaV_c)=0
\label{super}
\end{equation}

\begin{figure}
\vspace{1cm}
\includegraphics[width=11cm]{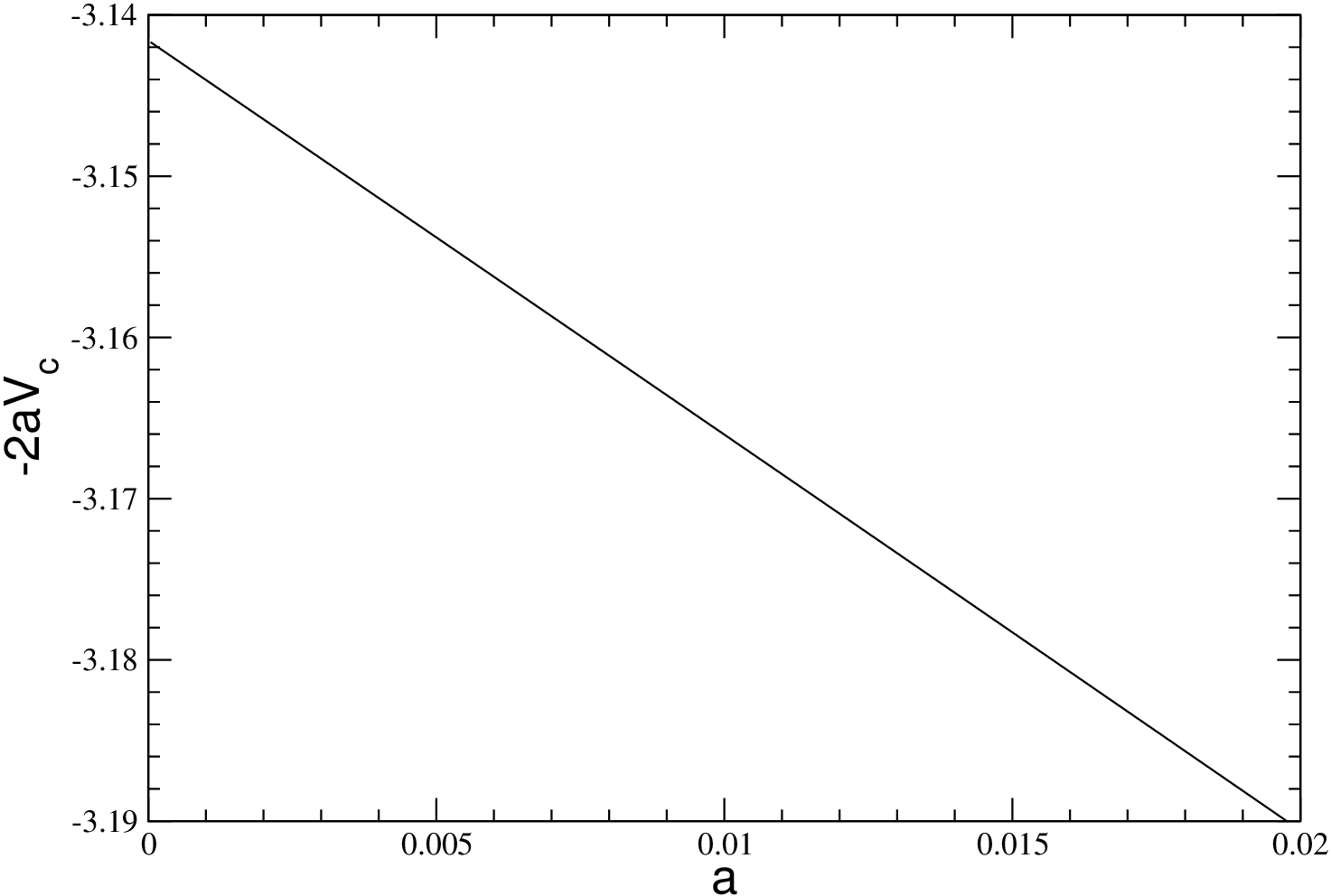}
\caption{\label{fig:Fig1} Plot of the supercritical potential
strength $V$ ($E=-m$) against the shape $a$.}
\end{figure}

As soon as the potential strength $V$ becomes larger than the
supercritical value $V_{c}$, the lowest bound state dives into the
energy continuum becoming a resonance. Fig. \ref{fig:Fig1} shows
the dependence of the supercritical potential $V_{c}$ on the shape
$a$. As $a\rightarrow 0$ we have that  $-2aV_{c}$ approaches $\pi$,
showing that the Dirac equation in the presence of a vectorial delta
interaction of strength $g_{v}$ becomes supercritical as $g_{v}=\pi$
\cite{FD:89}.

\begin{figure}
\vspace{1cm}
\includegraphics[width=11cm]{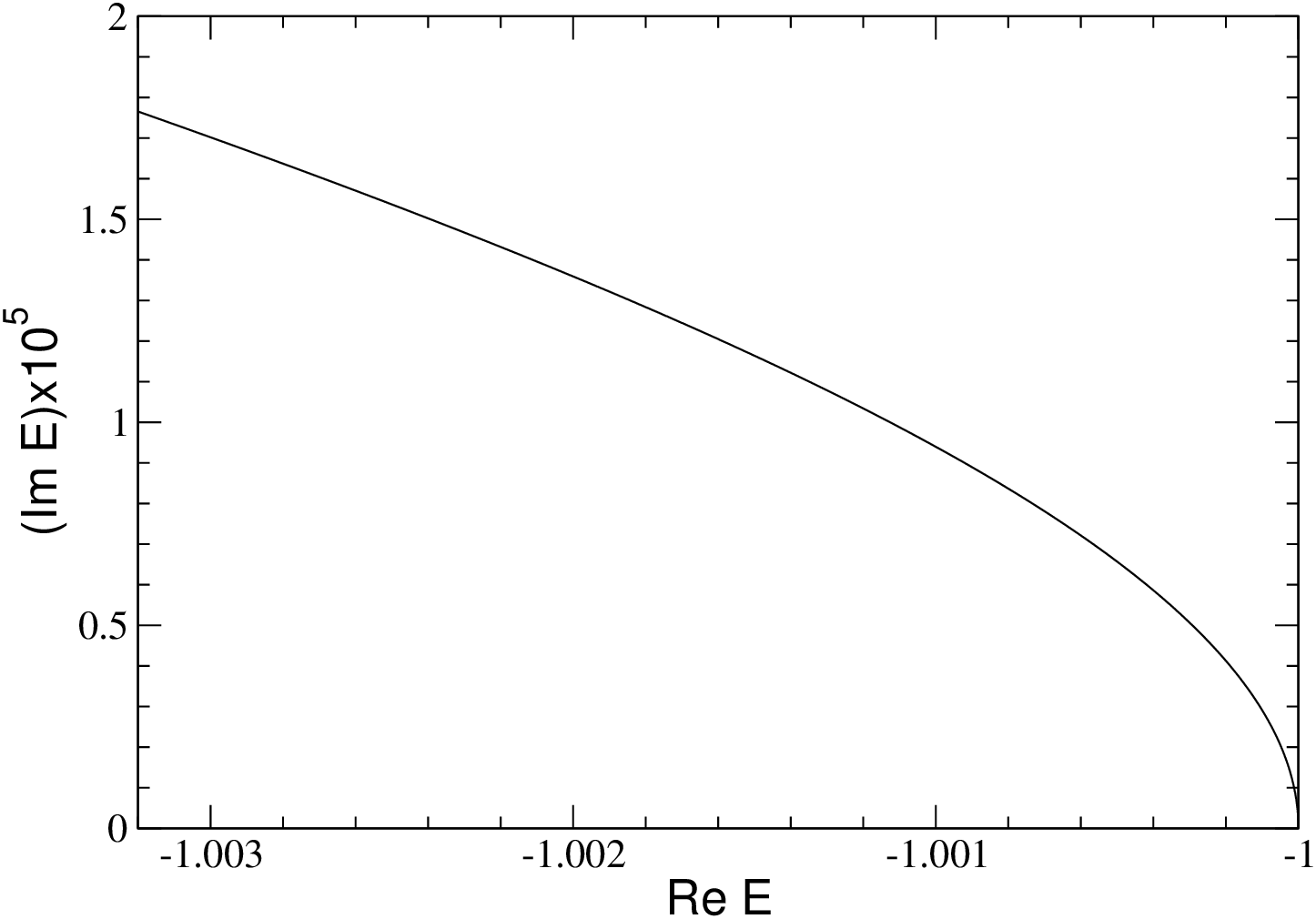}
\caption{\label{fig:Fig2} Resonant energy spectrum for $a=1.57$ with
$3.145533<V<3.15$}
\end{figure}

Fig.  \ref{fig:Fig2} shows, for a given value of the shape
parameter $a$, the dependence of the resonant energy of the lowest
bound state on the potential strength $V$. The curve of the
imaginary part as a function of the real part of the resonant energy
shows a concavity that decreases as the potential $V$ increases.

\section{Transmission amplitude and phase shifts}

In order to study resonant states associated with Dirac particles in
a cusp potential, we proceed to discuss the scattering process with
the help of the Jost functions \cite{Galindo,Newton}. Resonances can
be identified as poles of the scattering matrix $S$
\cite{Galindo,Goldberger,Newton}. Since the components of the
scattering matrix as well as the reflection and transmission
coefficients $R$ and $T$ can be expressed in terms of the reflection
and transmission amplitudes, we proceed to compute $r$ and $t$ with
the help of the incoming $\phi^{\pm}(x)$ and outgoing $\psi^{\pm}$
Jost solutions, where plus and minus indicate the asymptotic right
and left solutions.

Since the cusp potential (\ref{A1}) decays faster than $1/x^2$ for
large values of $x$, there are solutions of the Dirac equation
(\ref{Dirac}) that behave asymptotically as free traveling plane
wave spinors. Those solutions exhibiting this asymptotic behavior
are called Jost solutions \cite{Newton}. The one-dimensional Dirac
equation possesses  two outgoing $\psi^{\pm}$ and two incoming
$\varphi^{\pm}$ Jost solutions \cite{Thaller,Dreizler}

Since the Jost solutions $\psi_{\pm}(E,x)$ and $\varphi_{\pm}(E,x)$,
are linearly independent, we can write  regular solutions
$\Psi^{(-)}\left(E,x\right)$, $\Psi^{(+)}\left(E,x\right)$ of the
Dirac  equation as \cite{Thaller}
\begin{eqnarray}\label{expansion-sol-jost}
\fl
\Psi^{(\pm)}\left(E,x\right)=-\frac{1}{W\left(\psi_{\pm},\varphi_{\pm}\right)(E)}[W\left(\Psi^{(\pm)}\left(E,x\right)
,\psi_{\pm}\left(E,x\right)\right)\varphi_{\pm}\left(E,x\right)=\\
\nonumber
-W\left(\Psi^{(\pm)}\left(E,x\right),\varphi_{\pm}\left(E,x\right)\right)\psi_{\pm}\left(E,x\right)]\\
\nonumber
=-\frac{1}{W\left(\psi_{\pm},\varphi_{\pm}\right)(E)}\left[f^{(\pm)}_{+}\left(E\right)\varphi_{\pm}\left(E,x\right)-f^{(\pm)}_{-}
\left(E\right)\psi_{\pm}\left(E,x\right)\right]
\end{eqnarray}
where  the  Wronskian  $W\left(\psi(x),\varphi(x)\right)$ of the two
two-component solutions $\psi(x)$ and $\varphi(x)$ of the Dirac
equation $\psi(x)=\left(\psi_{1}(x),\psi_{2}(x)\right)^{t}$ and
$\varphi(x)=\left(\varphi_{1}(x),\varphi_{2}(x)\right)^{t}$ is
defined as \cite{Calogero,Kennedy2}
\begin{equation}\label{wronskiano}
W\left(\psi(x),\varphi(x)\right)\equiv\det\left(%
\begin{array}{cc}
\psi_{1}(x) & \varphi_{1}(x)  \\
\psi_{2}(x) & \varphi_{2}(x) \\
\end{array}%
\right) =\psi_{1}(x)\varphi_{2}(x)-\psi_{2}(x)\varphi_{1}(x)
\end{equation}
It is easy to see that if $W(\psi,\varphi)\neq 0$ we have that
$\psi(x)$ and $\varphi(x)$ are linearly independent,
$f_{-}^{(\pm)}\left(E\right)$, $f_{+}^{(\pm)}\left(E\right)$ are in
Eq (\ref{expansion-sol-jost}), analogous to the non-relativistic
case, the Jost functions \cite{Newton}.

We make use of the Jost solutions in order to describe a scattering
process. The  solution $\psi_{+}$ should be a linear combination of
the solutions $\psi_{-}$ and  $\varphi_{-}$, that is,
\begin{equation}\label{scatt-jost-1}
t^{\pm}(E)\,\psi_{\pm}(E,x)=\varphi_{\mp}(E,x)+r^{\mp}(E)\,
\psi_{\mp}(E,x)
\end{equation}
Since the cusp  potential (\ref{A1}) is even,  using the invariance
under parity \cite{Greiner} and the representation (\ref{rep}) for
the gamma matrices we have that
\begin{equation}
\psi_{\pm}(E,x)=-i\sigma_{y}\psi_{\mp}(E,-x), \
\varphi_{\pm}(E,x)=-i\sigma_{y}\psi_{\mp}(E,-x) \label{p1}
\end{equation}
The parity conditions (\ref{p1}) imply that the left and right
transmission and reflection amplitudes are identical. The
transmission amplitude $t(E)=t^{\pm}(E)$  can be expressed as
\begin{equation}
\label{coefs-trans} t(E)\equiv
t^{\pm}(E)=-\frac{W\left(\psi_{+},\varphi_{+}\right)(E)}
{W\left(\psi_{-},\psi_{+}\right)(E)}
\end{equation}
analogously, we have that the reflection amplitude $r(E)=r^{\mp}(E)$
reads
\begin{equation}\label{coefs-reflex}
r(E)\equiv
r^{\mp}(E)=\frac{W\left(\psi_{+},\varphi_{-}\right)(E)}{W\left(\psi_{-},\psi_{+}\right)(E)},
\end{equation}
The incoming Jost solution $\varphi_{-}$ for the one dimensional
Dirac equation (\ref{Dirac}) takes the form
\begin{equation}\label{varphi-menos-cuspide}
\varphi_{-}(\bar{y})=\frac{m}{(-2iaeV)^{\mu}\sqrt{2E(E-\sqrt{E^{2}-m^{2}})}}\left(
\begin{array}{c}
\frac{\frac{1}{2}+k-\mu}{ma}\,\bar{y}^{-1/2}M_{k,\mu }(\bar{y})\\
\bar{y}^{-1/2}M_{k+1,\mu }(\bar{y}),
\end{array}
\right)
\end{equation}
a spinor that exhibits the following asymptotic behavior
\begin{equation}\label{asint-varphi-menos-cuspide}
\varphi_{-}(x)\rightarrow
\frac{m}{\sqrt{2E(E-\sqrt{E^{2}-m^{2}})}}\left(
\begin{array}{c}
\frac{\frac{1}{2}+k-\mu}{ma} \\
1
\end{array}
\right) \exp (ix\sqrt{E^{2}-m^{2}}.)
\end{equation}
We can see that the solution (\ref{varphi-menos-cuspide}) behaves
asymptotically as a traveling to the right plane wave solution to
the free Dirac equation. The outgoing Jost solution $\psi_{-}$ can
be written as
\begin{equation}\label{psi-menos-cuspide}
\psi_{-}(\bar{y})=\frac{m}{(-2iaeV)^{-\mu}\sqrt{2E(E+\sqrt{E^{2}-m^{2}})}}\left(
\begin{array}{c}
\frac{\frac{1}{2}+k+\mu
}{ma}\bar{y}^{-1/2}M_{k,-\mu }(\bar{y})\\
\bar{y}^{-1/2}M_{k+1,-\mu }(\bar{y}),
\end{array}
\right)
\end{equation}
such a spinor exhibits the following asymptotic behavior as
$x\rightarrow -\infty$
\begin{equation}\label{asint-psi-menos-cuspide}
\psi
_{-}(x)\rightarrow\frac{m}{\sqrt{2E(E+\sqrt{E^{2}-m^{2}})}}\left(
\begin{array}{c}
\frac{\frac{1}{2}+k+\mu
}{ma} \\
1
\end{array}
\right) \exp (-ix\sqrt{E^{2}-m^{2}}).
\end{equation}
This asymptotic form shows that the Jost solution
(\ref{psi-menos-cuspide}) asymptotically behaves like a traveling
left solution wave to the  free Dirac equation. The right outgoing
Jost function $\psi_{+}$ is

\begin{equation} \label{psi-mas-cuspide}
\fl \psi
_{+}(y)=\frac{m}{(-2iaeV)^{-\mu}\sqrt{2E(E-\sqrt{E^{2}-m^{2}})}}\left(
\begin{array}{c}
\frac{\frac{1}{2}+k-\mu
}{ma}y^{-1/2}M_{k+1,-\mu }(y)\\
y^{-1/2}M_{k,-\mu }(y)
\end{array}
\right),
\end{equation}
such a spinor exhibits the following asymptotic behavior.
\begin{equation}\label{asint-psi-mas-cuspide}
\psi _{+}(x)\rightarrow
\frac{m}{\sqrt{2E(E-\sqrt{E^{2}-m^{2}})}}\left(
\begin{array}{c}
\frac{\frac{1}{2}+k-\mu}{ma} \\
1
\end{array}
\right) \exp (ix\sqrt{E^{2}-m^{2}})
\end{equation}
Finally, we have that the incoming Jost solution $\varphi_{+}$ takes
the form
\begin{equation} \label{varphi-mas-cuspide}
\varphi
_{+}(y)=\frac{m}{(-2iaeV)^{\mu}\sqrt{2E(E+\sqrt{E^{2}-m^{2}})}}\left(
\begin{array}{c}
\frac{\frac{1}{2}+k+\mu }{ma}y^{-1/2}M_{k+1,\mu }(y) \\
y^{-1/2}M_{k,\mu }(y)
\end{array}
\right)
\end{equation}
whose asymptotic behavior is that of an incoming wave from the
right.
\begin{equation}\label{asint-varphi-mas-cuspide}
\varphi _{+}(x)\rightarrow
\frac{m}{\sqrt{2E(E+\sqrt{E^{2}-m^{2}})}}\left(
\begin{array}{c}
\frac{\frac{1}{2}+\mu +k}{ma} \\
1
\end{array}
\right) \exp (-ix\sqrt{E^{2}-m^{2}})
\end{equation}

With the help of  the Jost functions and using Eq.
(\ref{coefs-trans}), we have  that the transmission amplitude $t$
for the Dirac equation (\ref{Dirac}) takes the form

\begin{equation}\label{t-izq-cuspide}
\fl
t(E)=\frac{ma}{\frac{1}{2}+k+\mu}\frac{\frac{\frac{1}{2}+k+\mu}{ma}M_{k+1,\mu
}(-2iaeV)M_{k,-\mu}(-2iaeV)-\frac{\frac{1}{2}+k-\mu}{ma}M_{k+1,-\mu}(-2iaeV)M_{k,\mu}(-2iaeV)}{(-2iaeV)^{2\mu}(M_{k,-\mu}^{2}(-2iaeV)+
\frac{\left(\frac{1}{2}+k-\mu\right)^{2}}{m^{2}a^{2}}M_{k+1,-\mu}^{2}(-2iaeV))}
\end{equation}
Analogously, we have that the reflection amplitude $r(E)$ is
\begin{equation}\label{r-izq-cuspide}
\fl r(E)=i\frac{ma}{\frac{1}{2}+k+\mu}\frac{M_{k+1,\mu
}(-2iaeV)M_{k+1,-\mu}(-2iaeV)-M_{k,-\mu}(-2iaeV)M_{k,\mu}(-2iaeV)}{(-2iaeV)^{2\mu}(M_{k,-\mu}^{2}(-2iaeV)+
\frac{\left(\frac{1}{2}+k-\mu\right)^{2}}{m^{2}a^{2}}M_{k+1,-\mu}^{2}(-2iaeV))}
\end{equation}

\begin{figure}
\vspace{1cm}
\includegraphics[width=11cm]{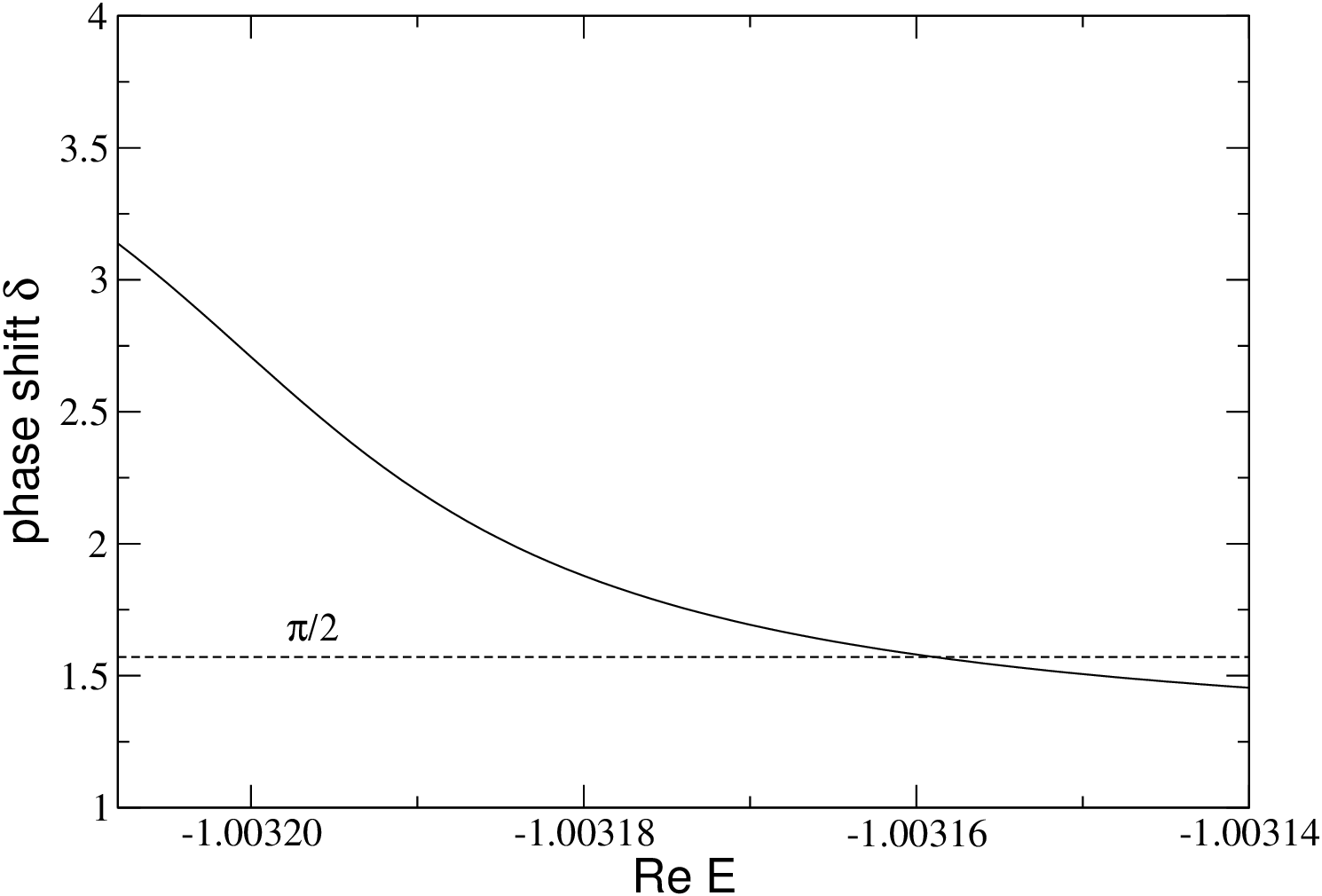}
\caption{\label{fig:Fig7} Phase shift $\delta$ as function of the
real part of the energy $E$ for $V=3.15$, $a=1.57$. Notice that the
phase shift $\delta$ of the resonance value $\Re E=-1.0032$ is
larger than $\pi/2$}
\end{figure}

\begin{figure}
\vspace{1cm}
\includegraphics[width=11cm]{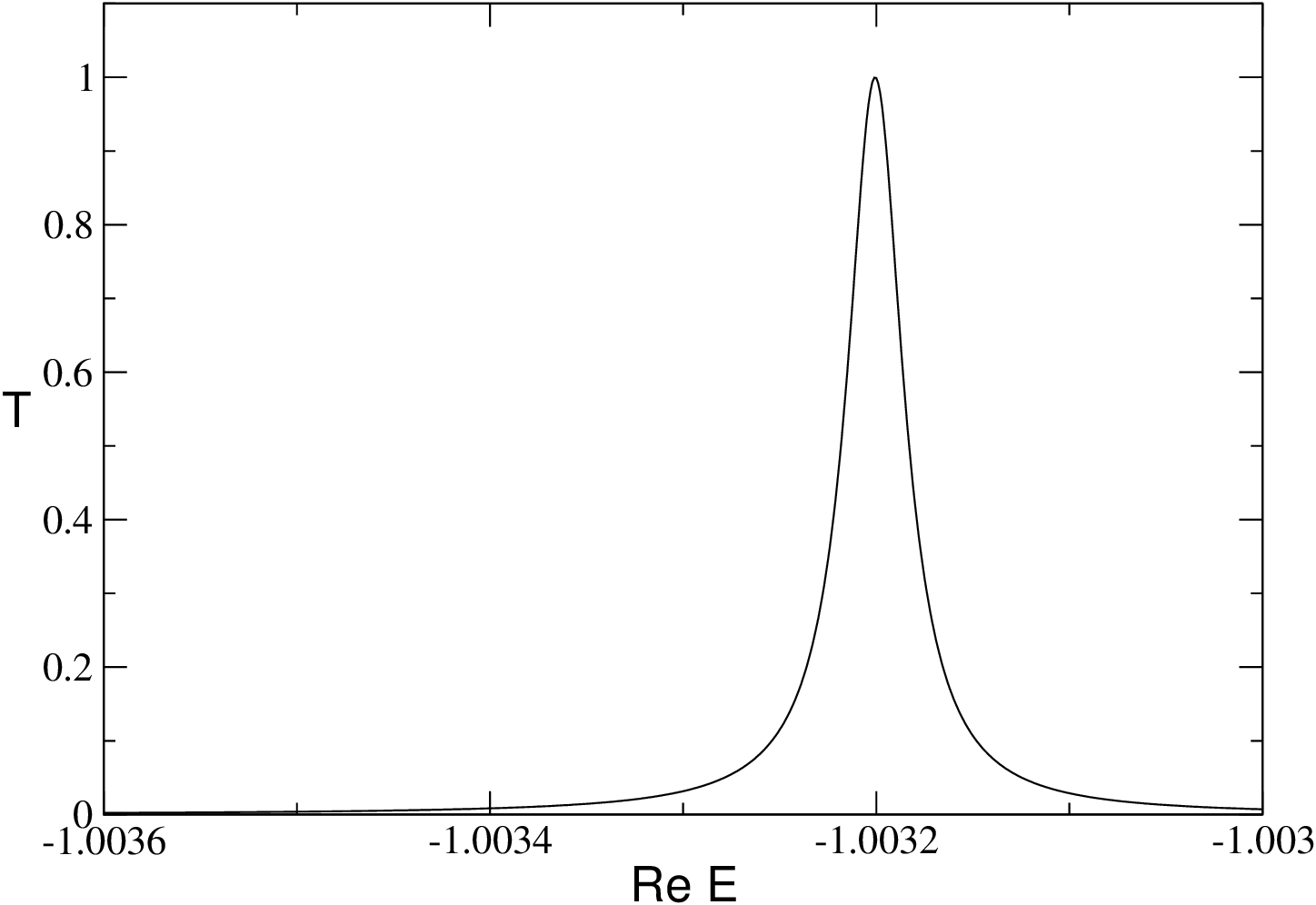}
\caption{\label{fig:Fig5} Transmission coefficient for $V=3.15$,
$a=1.57$. A maximum corresponding to a resonance is reached for $\Re
E=-1.0032$}
\end{figure}

\begin{figure}
\vspace{1cm}
\includegraphics[width=11cm]{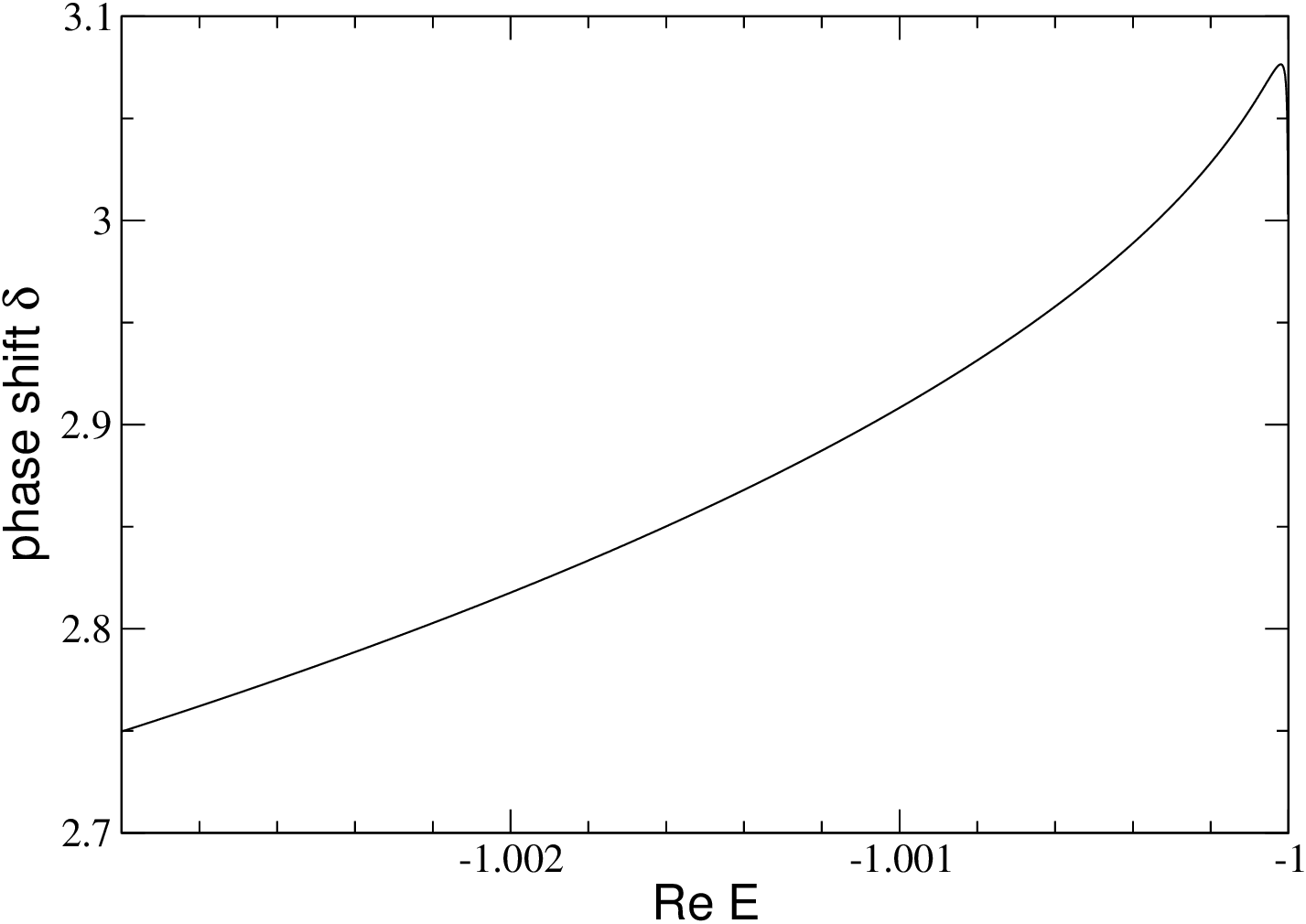}
\caption{\label{fig:Fig9} Phase shift $\delta$ as a function of the
real part of the resonant energy for  $3.145533<V<3.15$ with
$a=1.57$}
\end{figure}

It is not difficult to see that the components $t(E)$ and $r(E)$ of
the scattering matrix $S$  have the same poles, they are the roots
of the Jost function $f_{+}(E)$

\begin{equation}\label{func-jost-cuspide}
\fl
f_{+}(E)=W\left(\psi_{-},\psi_{+}\right)(E)=M_{k,-\mu}^{2}(-2iaeV)+
\frac{\left(\frac{1}{2}+k-\mu\right)^{2}}{m^{2}a^{2}}M_{k+1,-\mu}^{2}(-2iaeV)
\end{equation}
It is worth noticing that the equation $f_{+}(E)=0$ coincides with
the resonant energy condition given by Eq. (\ref{energy}).

When the transmission coefficient reaches the unit value, that is
for $E=\Re\,E_{res}$, where $E_{res}$ is the resonance energy, the
phase shift $\delta_{res}(E)$ reaches the value $\pi/2$. In the
vicinity of the resonant value, provided that the Breit-Wigner
formula \cite{Newton} is valid, the resonance energy is a simple
zero of the Jost function $f^{(\pm)}_{+}(E)$.  In the neighborhood
of $E=\Re\,E_{res}$ we have

\begin{equation}
\label{apro}
f^{(\pm)}_{+}(E)\approx\left(\frac{df^{(\pm)}_{+}}{dE}\right)_{E=E_{res}}\left(E-E_{res}\right).
\end{equation}
If $\Im\,E_{res}$ is sufficiently close to the real axis for the
approximation (\ref{apro}) to be applicable in a real neighborhood
of $E=\Re\,E_{res}$ we will have
\begin{equation}
\delta(E)=-arg(f_{+}(E))_{E=E_{res}}\approx\delta_{bg}(E)+\delta_{res}(E)
\end{equation}
hence
\begin{equation}
\delta_{res}(E)\approx\delta(E)-\delta_{bg}(E),
\end{equation}
where $\delta_{bg}=-arg\left(\frac{df_{+}}{dE}\right)_{E=E_{res}}$
is the background phase shift. Fig  \ref{fig:Fig7} shows the
dependence of the phase shift $\delta$ on the real part of the
energy  $E$. For the resonance value $\Re E=-1.0032$ we have that
$\delta=2.62769$, and $\delta_{bg}=1.05706$, obtaining in this way
$\delta_{res}=1.57063\approx \pi/2$.

It is worth mentioning that as the lowest-bound energy  state reaches
the supercritical value $E=-m$,
the phase shift $\delta$   takes the value $\delta=\pi/2$,  result that is in
agreement with the phase shift for
the one-dimensional Dirac equation in the presence of  a symmetric potential
well \cite{Greiner,Calogeracos}

The Wigner time $\tau$ permits one to estimate how long the bound
state lives before it degenerates into the negative continuum. From
Fig. \ref{fig:Fig5} and the solution to Eq. (\ref{energy}), we can
estimate the mean life of the resonant state in terms of the
half-width $\Gamma$ of the Lorentzian curve. For $\Re E=-1.0032$ we
have $\Im E=1.765 \times 10^{-5}$, corresponding to a Wigner time of
$\tau=\frac {d\delta}{dE}\approx \frac{2}{\Gamma}=5.667 \times
10^{5}$ expressed in natural units. Resonant states make the phase
shift surpass the value $\pi/2$ with a Wigner time delay $\tau>0$
Fig. \ref{fig:Fig9} shows the behavior of the phase shift $\delta$
as a function of the real part of the resonant energy, it can be
observed that the phase shift increases surpassing the value $\pi/2$
and after reaching a maximum, it monotonically decreases, showing in
this way a behavior that has been observed in square well case.
\cite{KD:04}

\section{Concluding remarks}

The relation (\ref{energy}) shows that the one dimensional cusp
potential (\ref{A1}) supports supercritical states and is strong
enough to produce resonant states. This result is non trivial and
interesting in view of the fact that the Dirac equation in a
one-dimensional vectorial dirac delta interaction does not exhibit
resonances \cite{FD:89}. The potential (\ref{A1}) does not exhibit a
square barrier limit, and for very small values of $a$ and constant
value of $2Va$ it can be regarded as a delta potential
\cite{Dominguezadame}, therefore the the relation (\ref{super}) also
holds for a delta potential. Eq. (\ref{A1}) represents a local
interaction whose support does not vanish anywhere and exhibits
supercritical resonant states.

We have computed the phase shift associated with the scattering of a
Dirac particle by the potential (\ref{A1}), and have shown that
resonant states make  the phase shift surpass the value $\pi/2$ with
a Wigner time-delay $\tau>0$. The asymptotic behavior of the cusp
potential has permitted us to derive the transmission and reflection
amplitudes $t(E)$ and $r(E)$ with the help of the Jost solutions.
This approach can be applied in the study of resonant states in the
presence of potentials not allowing exact solutions of the Dirac
equation in terms of special functions, this problem will be
discussed in a forthcoming publication.

\ack{We thank Dr. Ernesto Medina for reading and improving the
manuscript. One of the authors (VMV) wants to express his gratitude
to the Alexander von Humboldt Foundation for financial support.}

\end{document}